\documentclass[conference]{IEEEtran}
\IEEEoverridecommandlockouts

\ifCLASSINFOpdf
   \usepackage[pdftex]{graphicx}
\else
\fi
\usepackage{amsmath}
\usepackage{verbatim,subfloat,pifont}

\usepackage{array}

  \usepackage[caption=false,font=footnotesize]{subfig}
\usepackage{url}

\begin{document}
%
\title{Experimental and simulation study of irradiated silicon pad detectors for the CMS High Granularity Calorimeter}

\author{\IEEEauthorblockN{Timo Peltola, on behalf of the CMS collaboration}
\IEEEauthorblockA{Texas Tech University\\
Department of Physics and Astronomy, Lubbock, Texas 79409--1051\\
Email: timo.peltola@ttu.edu}}

\IEEEpubid{978-1-5386-2282-7/17/\$31.00~\copyright~2017 IEEE}

\maketitle

\begin{abstract}
\label{abstract}
The foreseen upgrade of the LHC to its high luminosity phase (HL-LHC), will maximize the physics potential of the facility. The upgrade is
expected to increase the instantaneous luminosity by a factor of 5 and deliver an integrated luminosity of 3000~fb$^{-1}$ after 10 years of
operation. As a result of the corresponding increase in radiation and pileup, the electromagnetic calorimetry in the CMS endcaps will sustain maximum
integrated doses of 1.5~MGy and neutron fluences above $10^{16}~\textrm{n}_\textrm{eq}\textrm{cm}^{-2}$, necessitating their replacement for HL-LHC
operation.

The CMS collaboration has decided to replace the existing endcap electromagnetic and hadronic calorimeters by a High Granularity Calorimeter (HGCAL) that will provide unprecedented information on electromagnetic and hadronic showers in the very high pileup of the HL-LHC. 
In order to employ Si detectors in HGCAL and to address the challenges brought by the intense radiation environment, an extensive R\&D program has
been initiated, comprising production of prototype sensors of various types, sizes and thicknesses, their qualification before and after
irradiation to the expected levels, and accompanying simulation studies. 

The ongoing investigation presented here includes measurements of current-voltage and capacitance-voltage characteristics, along with
predicted charge collection efficiences of the sensors 
irradiated to levels expected for the HGCAL at HL-LHC. 
The status of the study 
and the first 
results of the performance of neutron irradiated Si detectors, as well as their comparison 
with numerical simulations, 
are presented. 
%

\end{abstract}


%
\IEEEpeerreviewmaketitle

\section{Introduction}
\label{intro}
\IEEEPARstart{T}{he} endcap calorimeters in the Compact Muon Solenoid (CMS) experiment will be replaced, around 2024, by the High Granularity Calorimeter. 
The HGCAL
will be realized as a sampling calorimeter with 
52 layers of active material. The electromagnetic section and the high-radiation region of the
hadronic section will use hexagonal silicon sensors as active material. The low-radiation regions of the hadronic section will use plastic
scintillator tiles with on-tile silicon photomultipliers (SiPM). The silicon sensors will be divided into cells of $\sim0.5-1.0~\textrm{cm}^2$
and will have active thicknesses from 100 to 300 $\mu\textrm{m}$ depending on their pseudorapidity (thinner sensors at higher $\eta$) \cite{Magnan2017}.

Position sensitive silicon (Si) detectors have already been extensively employed in the tracking systems of High Energy
Physics experiments due to their outstanding performance and radiation hardness, as well as in some calorimeters. For example, the
CMS Preshower detector includes two layers of silicon sensors in each endcap, upstream of the main part of the electromagnetic calorimeter.
In order to employ Si detectors in HGCAL and to address the challenges
caused by the intense radiation environment, extensive measurements and simulation studies of Si pad detectors have been initiated by
the CMS collaboration.
\IEEEpubidadjcol

Essential information of the performance of an irradiated silicon detector is obtained by monitoring its charge collection efficiency (CCE).
From the evolution of the CCE as a function of fluence it is possible to directly observe the effect of the radiation induced defects on the
ability of the detector to collect charge carriers generated by traversing 
ionizing particles.  
By complementing observed CCE with leakage current and capacitance measurements, a comprehensive picture of the macroscopic effects from
the microscopic radiation-induced defects can be obtained.

Simulations are a vital tool for e.g. device structure optimization or predicting the electric fields and trapping in the silicon sensors.
When numerical simulations are able to verify experimental results they also gain predictive power, resulting in reduced time and cost budget
in detector design and testing. Technology Computer-Aided Design (TCAD) simulations of silicon strip sensors have expanded to cover both bulk
and surface properties after irradiation at HL-LHC levels, producing results that are converging with measurements \cite{Peltola2015r,Peltola2016}.

In the following we present the first results of the
radiation tolerance study 
of neutron irradiated 300, 200, and 100~$\mu\textrm{m}$ 
active thickness p-on-n and n-on-p silicon pad detectors. 
The measured leakage currents, and 
full depletion voltages ($V_\textrm{fd}$) 
are first compared with the TCAD simulations, and then the simulation results for expected CCE are presented. 

%
%

\section{Measurements and simulations of neutron irradiated pad detectors}
\label{measSim}
\subsection{Samples and experimental setup}
Since the fluence in the HGCAL 
will be dominated by 
neutrons 
\cite{Curras2017},
an irradiation campaign of 16 
n-on-p and p-on-n samples, consisting of test diodes from HGCAL 
prototype hexagonal silicon sensor wafers, has been initiated at Rhode Island reactor\footnote{http://www.rinsc.ri.gov/}. 

As shown in table~\ref{table_fs}, ten of these samples have at the moment been both irradiated and capacitance-voltage ($CV$) and current-voltage ($IV$) characterized. For the $CV/IV$ characterization a novel probestation has been constructed at Texas Tech University (TTU) that provides cooling in dry environment and bias voltages up to 2.2 kV for the measurements of heavily irradiated Si sensors. In the immediate future the measurement facility at TTU will include a Transient Current Technique (TCT) setup to complete the characterization of the bulk properties of irradiated Si sensors by providing CCE data. 
The construction of the TCT setup is in progress and 
it is already operational for room temperature (RT) measurements. 
\begin{table}[!t]
\renewcommand{\arraystretch}{1.3}
\caption{Fluences determined from the irradiation times and the 1-MeV n$_\textrm{eq}$ flux at Rhode Island reactor, and the irradiated samples of varied polarities and active thicknesses.}
\label{table_fs}
\centering
\begin{tabular}{|c||c|c|c|c|}
\hline
\multicolumn{1}{|c|}{{\bf Fluence}} & \multicolumn{2}{c|}{{\bf p-on-n}} & \multicolumn{2}{c|}{{\bf n-on-p}}\\
\cline{2-5}
\multicolumn{1}{|c|}{[n$_\textrm{eq}$cm$^\textrm{-2}$]} & \multicolumn{1}{c|}{300 $\mu\textrm{m}$} & \multicolumn{1}{c|}{120 $\mu\textrm{m}$} & \multicolumn{1}{c|}{300 $\mu\textrm{m}$} & \multicolumn{1}{c|}{200 $\mu\textrm{m}$}\\
\hline
$(1.5\pm0.3)\times10^{14}$ & 1 & & 1 & \\
\hline
$(5.0\pm1.0)\times10^{14}$ & 1 & & 1 & \\
\hline
$(7.4\pm1.6)\times10^{14}$ & 1 & & 2 & \\
\hline
$(1.5\pm0.3)\times10^{15}$ & & & & 1\\
\hline
$(3.8\pm0.8)\times10^{15}$ & & 1 & & 1\\
\hline
\end{tabular}
\end{table}
%
\subsection{Leakage currents and extracted fluences}
\label{secLCV}
\begin{figure}[!t]
\centering
\includegraphics[width=3.5in]{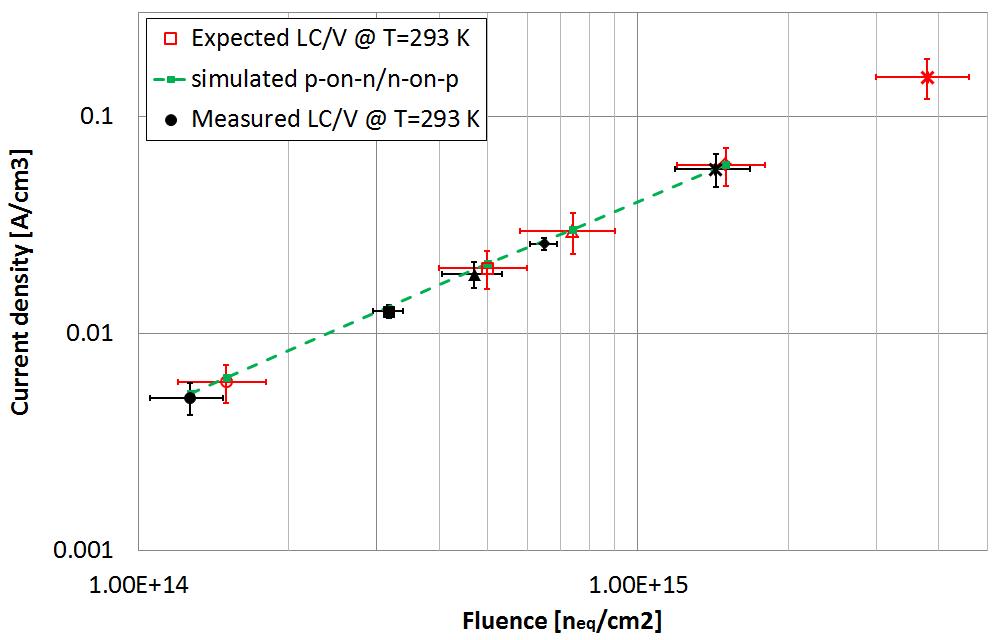}
\caption{Measured, simulated and expected from eq. \ref{eq1} current densities as a function of $\Phi_\textrm{eq}$.}
\label{fig_LCV}
\end{figure}

The experimentally determined leakage current densities shown in figure~\ref{fig_LCV} were reached by measuring the leakage current values at $V_\textrm{fd}$ that were extracted from $CV$-measurements.

Since current-related damage rate $\alpha$ is defined by \cite{Mollphd1999} 
\begin{equation}\label{eq1}
\Delta{I}/\textrm{Vol.}=\alpha\times\Phi_\textrm{eq},
\end{equation}
where $\Delta{I}$ is the change in the leakage current due to irradiation, $\textrm{Vol.}$ is the active volume of the detector and $\Phi_\textrm{eq}$ is the 1-MeV neutron equivalent fluence, it is possible to determine $\Delta{I}/\textrm{Vol.}$ extracted $\Phi_\textrm{eq}$ by using $\alpha(\textrm{293 K})=4.0\times10^{-17}~\textrm{A/cm}$ \cite{Moll1999,Lindstrom2003} for the measurements in RT. When the measured and expected $\Delta{I}/\textrm{Vol.}$ are plotted as a function of leakage current extracted and nominal values of $\Phi_\textrm{eq}$ from table~\ref{table_fs}, respectively, it can be seen from figure~\ref{fig_LCV} that the two agree only for the lowest fluence, while $\Delta\Phi_\textrm{eq}$ increases to about factor 2 for the highest fluences. The increase of $\Delta\Phi_\textrm{eq}$ with fluence could be the result of unwanted annealing during the processes of irradiation (irradiation times varied from 6 minutes to about 2.5 hours), overnight shipment, storage and testing (about 30 minutes out of the cold storage), and remains under investigation.

For the simulated $\Delta{I}/\textrm{Vol.}$ in figure~\ref{fig_LCV} the parameters of the modelled structures (bulk doping, active thickness, backplane deep-diffusion doping profile and charge carrier trapping times) were tuned to reproduce the measured $CV/IV$ results before irradiation. Then the neutron irradiation was modelled by CMS neutron defect model, which is validated for $\Phi_\textrm{eq}=1\times10^{14}\sim1\times10^{15}~\textrm{cm}^{-2}$ at fixed $T=253~\textrm{K}$ \cite{Eber2013}. As can be seen from the dashed green curve in figure~\ref{fig_LCV}, the simulation reproduces both measured and expected $\Delta{I}/\textrm{Vol.}$ when different fluence values are used as an input and the leakage currents are scaled to $T=293~\textrm{K}$ (The highest nominal fluence value is out of the validated range of neutron defect model and was not simulated).
%

All simulations in this study were carried out using the Synopsys Sentaurus\footnote{http://www.synopsys.com} finite-element TCAD software framework.
%
\subsection{Full depletion voltages}
\label{secVfd}
\begin{figure}[!t]
\centering
\includegraphics[width=3.4in]{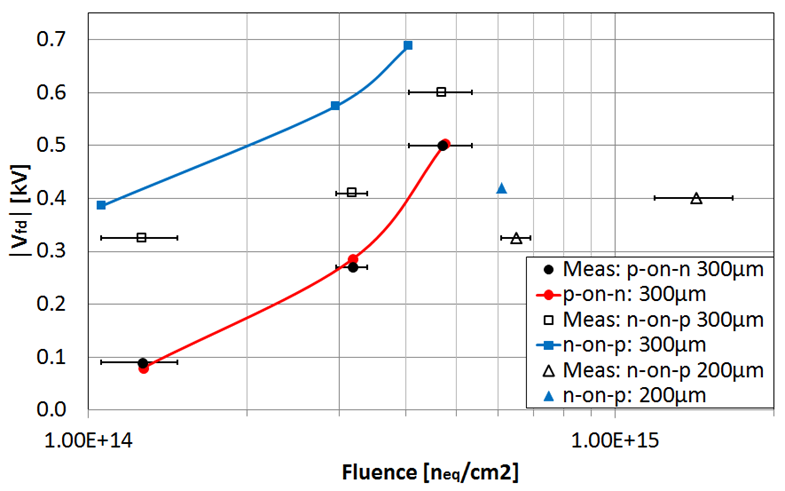}
\caption{Measured and simulated $V_\textrm{fd}$ as a function of leakage current extracted $\Phi_\textrm{eq}$.}
\label{fig_Vfd}
\end{figure}

Measured and simulated fluence evolution of $V_\textrm{fd}$ is presented in figure~\ref{fig_Vfd}. Since $\Delta{V_\textrm{fd}}$ for nominal $\Phi_\textrm{eq}$ values from table~\ref{table_fs} between measured and simulated results was in the range of several hundred of volts for all except the lowest fluence value, only the leakage current extracted $\Phi_\textrm{eq}$ values are considered in figure~\ref{fig_Vfd}.
%

Results show that when leakage current extracted $\Phi_\textrm{eq}$ is used as an input for the simulation that uses frequences in the range of the $CV$-measurement, the simulated $V_\textrm{fd}$($\Phi_\textrm{eq}$) matches the measurement for 300 $\mu\textrm{m}$ thick p-on-n sensors, while the simulated results for 300 and 200 $\mu\textrm{m}$ thick n-on-p sensors are within $(100\pm50)~\textrm{V}$ from measured throughout the $\Phi_\textrm{eq}$ range.

Thus, when the simulated leakage currents and $V_\textrm{fd}$ are compared with the measured, the results suggest that the leakage current extracted $\Phi_\textrm{eq}$ values reflect closely the effective fluences received by the samples. 
%
%
%
%
%
\subsection{Expected charge collection efficiences}
\begin{figure}[!t]
\centering
\includegraphics[width=3.4in]{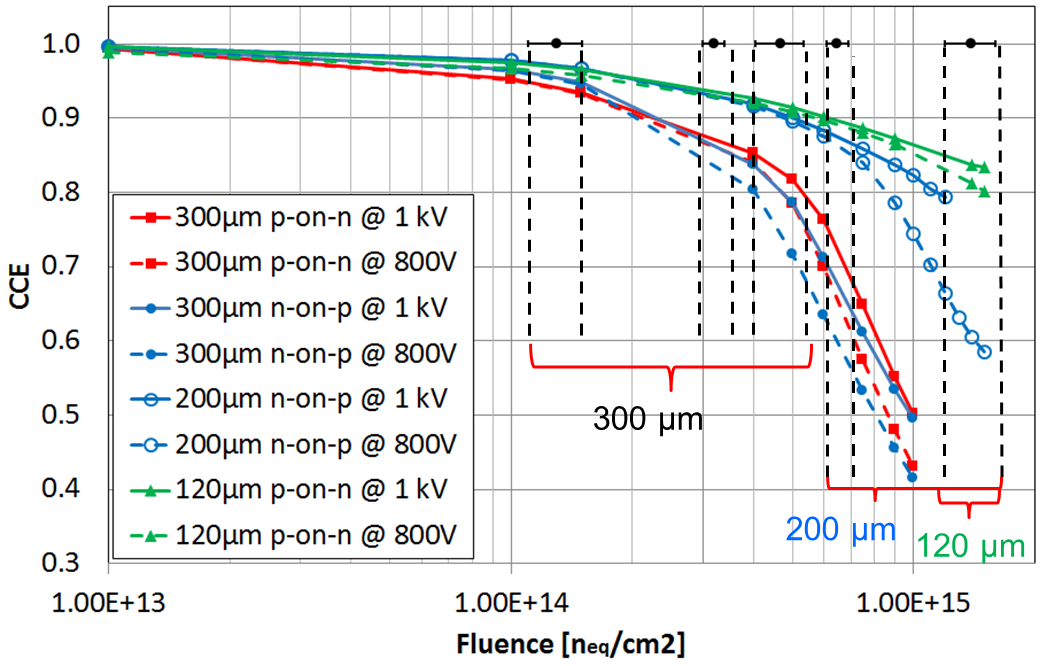}
\caption{Simulated charge collection efficiences as a function of $\Phi_\textrm{eq}$ for 800 V and 1 kV bias voltages. The leakage current extracted fluence regions for the three sensor thicknesses are indicated by black dashed lines to guide the eye.}
\label{fig_CCE}
\end{figure}

The agreement between the data and simulation in sections~\ref{secLCV} and~\ref{secVfd} gives confidence in using TCAD to make CCE predictions.
%
%
The 
results for the samples in table~\ref{table_fs} at bias voltages 800 V and 1 kV are presented in figure~\ref{fig_CCE}. The charge collection efficiency is determined as a ratio CCE = CC(irradiated)/CC(non-irradiated), where CC(irradiated) is the collected charge from minimum ionizing particle (MIP) injection at $T=253~\textrm{K}$ and given voltage, and CC(non-irradiated) is the collected charge from MIP-injection at RT and 500 V.
\enlargethispage{-1.00in}

%
%
%
%
%
%
In the fluence range indicated in figure~\ref{fig_CCE} for 300 $\mu\textrm{m}$ thick sensors, 
the p-on-n sensor is fully depleted at given voltages due to space charge sign inversion (SCSI) while equal thickness n-on-p is not. This is reflected in higher CCE for p-on-n sensor as well as 
CCE(n-on-p) benefiting more 
of the voltage increase from 800 V to 1 kV.

As expected, the thinner sensors display superior CCE to 300 $\mu\textrm{m}$ thick sensors while not close to their respective $\Phi_\textrm{max}$ due to the limited range of the neutron defect model.

\section{Summary, Conclusions and Outlook}
\label{Conclusion}
HGCAL Si pad detector irradiation campaign is underway at Rhode Island 
reactor.
For the characterization of the irradiated sensors a novel $CV/IV$-probestation has been built and is operational at TTU. 
%
A TCT-setup enabling CCE measurements 
of the sensors will be included in the immediate future. 

When the TCAD simulation is tuned with input from $CV/IV$-measurements before irradiation and implements leakage current extracted fluences within neutron defect model,  
measured $V_\textrm{fd}$ and leakage currents are closely reproduced.
Results for the two polarities of 300 $\mu\textrm{m}$ active thickness ($\Phi_\textrm{max}\approx6\times10^{14}~\textrm{n}_\textrm{eq}\textrm{cm}^{-2}$) sensors 
indicate that the 
p-on-n sensor configuration has lower $V_\textrm{fd}$ and higher CCE prediction for the expected fluence range 
than the n-on-p sensor. Additionally, to be considered in sensor polarity evaluation for HGCAL, the p-on-n sensor is cheaper to manufacture. 
Drawback of p-on-n sensor is the shifting of the electric field maximum to the backplane due to 
SCSI for $\Phi\geq1\times10^{14}~\textrm{n}_\textrm{eq}\textrm{cm}^{-2}$.
%

The upcoming 
study will include 
CCE and rise time 
measurements with infrared-TCT (with MIP-like carrier generation) 
enabling comprehensive investigation between sensor polarities and all three active thicknesses.


\section*{Acknowledgment}

The author would like to thank the colleagues at Brown University for their invaluable help during the irradiation campaign.



\bibliographystyle{IEEEtran}
\bibliography{mybibfile}
%
%
%

\end{document}